\documentclass{Interspeech2024}
\usepackage{cite}
\usepackage{latexsym}
\usepackage{comment}

\def\x{{\mathbf x}}
\def\c{{\mathbf c}}
\def\U{{\mathbf u}}

\def\V{{\cal V}}
\def\H{H}
\def\h{\mathbf h}
\def\Y{Y}

\def\X{{X}}
\def\Z{Z}
\def\F{\mathcal{F}}

\def\dec{\mathrm{dec}}
\def\CTC{\mathrm{CTC}}
\def\ctc{\mathrm{ctc}}

\def\MLP{\mathrm{MLP}}
\def\ctc{\mathrm{ctc}}
\def\ctx{\mathrm{cxt}}

\def\b{\mathrm{b}}
\def\n{\mathrm{n}}

\interspeechcameraready

\title{Decoder-only Architecture for Streaming End-to-end Speech Recognition}

\name[affiliation={1}]{Emiru}{Tsunoo}
\name[affiliation={1}]{Hayato}{Futami}
\name[affiliation={1}]{Yosuke}{Kashiwagi}
\name[affiliation={2}]{Siddhant}{Arora}
\name[affiliation={2}]{Shinji}{Watanabe}

\address{
  $^1$Sony Group Corporation, Japan\\
  $^2$Carnegie Mellon University, USA}
\email{emiru.tsunoo@sony.com}


\begin{document}

\maketitle

\begin{abstract}
\begin{comment}
    Decoder-only language models (LMs) have been successfully adopted for speech-processing tasks including automatic speech recognition (ASR).
    Owing to the ample expressiveness of the LMs, they perform efficiently even with the smaller parameter size than the off the shelf pretrained large language models.
    This efficiency is a suitable characteristic for streaming applications of ASR.
    In this work, we propose to use a decoder-only architecture for blockwise streaming ASR.
    In our approach, speech features are compressed using CTC greedy output as prompts to the LM.
    These prompts are sequentially generated from blockwise processing of the conformer-based speech subnetwork.
    The speech subnetwork introduces context embedding that represents past information, which is also provided to the decoder as additional prompts.
    The decoder estimates the output tokens promptly at each block.
    %To this end, we also propose a training scheme using random prefix prompts to make the model robust to the streaming inference where the prompts are limited.
    To this end, we also propose a novel training scheme using random-length prefix prompts to make the model robust to the truncated prompts caused by blockwise processing.
    An experimental comparison shows that our proposal decoder-only streaming ASR improves inference speed by almost a factor of 2 without significant accuracy degradation, or slight improvements by reducing word error rates of LibriSpeech test-other set by 0.7\% absolute.
\end{comment}
Decoder-only language models (LMs) have been successfully adopted for speech-processing tasks including automatic speech recognition (ASR). The LMs have ample expressiveness and perform efficiently. This efficiency is a suitable characteristic for streaming applications of ASR. In this work, we propose to use a decoder-only architecture for blockwise streaming ASR. In our approach, speech features are compressed using CTC output and context embedding using blockwise speech subnetwork, and are sequentially provided as prompts to the decoder. The decoder estimates the output tokens promptly at each block. To this end, we also propose a novel training scheme using random-length prefix prompts to make the model robust to the truncated prompts caused by blockwise processing. An experimental comparison shows that our proposed decoder-only streaming ASR achieves 8\% relative word error rate reduction in the LibriSpeech test-other set while being twice as fast as the baseline model.
\end{abstract}

\section{Introduction}
Streaming-style or online automatic speech recognition (ASR) is required in many real-world use cases, such as real-time transcription of broadcast contents.
Many end-to-end approaches has established for such an application, including blockwise processing encoders \cite{povey18,dong19, tsunoo19, shi2021emformer} combined with connectionist temporal classification (CTC) \cite{graves06, miao15, amodei16, arora2023semi} or transducers \cite{graves13rnnt, rao17,zhang2020transformer}.
Some studies use additional modules to align the sequential process of encoders with the attention-based transformer decoders \cite{moritz19, miao2020,Li23}, and the decoders are also used for score fusion or rescoring in the streaming scenario \cite{sainath2019two, zhou2021phoneme, yao2021wenet, tsunoo23_interspeech,botros2023lego}.
Although the attention decoders have a strong ability to model label sequences, the source--target attention layer is computationally costly because it attends on all the encoded features, which is generally longer than the label sequence.

Inspired by the recent advancement of decoder-only language models (LMs) such as GPT-3 \cite{brown2020language} and PaLM \cite{chowdhery2022palm}, the decoder-only LMs are adapted for speech-processing tasks \cite{chang2023speechprompt,zhang2023speechgpt,rubenstein2023audiopalm,fathullah2023prompting,wu2023decoder, arora23_interspeech, maiti2023voxtlm}.
Instead of using source--target attention layer, to bridge the audio--text modalities, audio information is alternately injected into the LLMs as prompts, which are discrete audio units  \cite{chang2023speechprompt, zhang2023speechgpt, rubenstein2023audiopalm} or continuous representations injected directly into the linguistic embedding space \cite{fathullah2023prompting, wu2023decoder}.
We previously demonstrated that a relatively small decoder trained with text-only data can achieve not only strong ASR performance but also computational efficiency \cite{tsunoo2023decoder}.
Such a characteristic is favorable for the streaming scenario.
However, there have not been any investigations in deploying decoder-only models for streaming ASR.

This study aims to use powerful yet efficient decoder-only architecture for blockwise streaming ASR.
Speech utterances are processed in a blockwise conformer-based speech subnetwork, and each block produces prompts that represent acoustic information.
The prompts are considerably compressed by removing unnecessary frames with the auxiliary CTC greedy search.
The speech subnetwork introduces context embedding that is inherited from previous blocks and represents past context information.
This context embedding is also provided to the decoder as additional prompts.
The decoder-only architecture estimates transcriptions immediately after the chunk of prompts is given.
% However, while training the model can use prompts corresponding to the entire utterance, the prompts are limited in inference due to the blockwise processing.

During training, however, it is difficult to prepare a partial transcription corresponding to the chunk of prompts without accurate forced alignment.
% Although on naive solution is to train a model with the entire transcription, even with partial prompts.
Although Sharma {\it et al.} train the blockwise model using the entire transcription without alignments by introducing chain rule of probability \cite{sharma23_interspeech}, none has explored the prompt alignment problem yet to the best of our knowledge.
Thus, we also propose a new training method that mitigates this training--inference mismatch.
To simulate the blockwise inference scenario, we select a random-length prefix of prompts to train the decoder.
%Experimentally we confirmed that our proposed usage of decoder-only architecture in streaming system performs better accuracy with a lower real-time factor (RTF), e.g., XX\% word error rate (WER) improvement on Librispeech test-other set with XX\% less computational cost.

The main contributions of this study are as follows:
\begin{itemize}
\item To the best of our knowledge, we are first to propose combining a decoder-only architecture with a blockwise speech subnetwork.
\item We propose to mitigate the mismatch between streaming inference and batch training by randomly selecting a prefix of prompts for training.
\item The proposed approach achieves higher ASR accuracy than streaming CTC/transducer models with only a small increase in computational cost, and even higher accuracy than ordinary encoder--decoder approach with faster inference.
\end{itemize}

\section{Streamable decoder-only architecture}
\label{sec:deconly}

\subsection{Decoder-only architecture for ASR}
\label{ssec:deconly}
%Conventionally, in ASR tasks, attention-based transformer decoders with source--target layers are used \cite{karita19}.
%Recently, however, autoregressive decoder-only LMs are also effectively adapted to ASR tasks \cite{rubenstein2023audiopalm, fathullah2023prompting, tsunoo2023decoder, wu2023decoder}.
% A transformer decoder can be fused with CTC for the better inference \cite{watanabe17}.
%To leverage both expressiveness and computational efficiency, we adopt the decoder-only architecture. 

\begin{figure}[t]
  %\centering
  \hspace{-0.5cm}
  \includegraphics[width=1.1\linewidth]{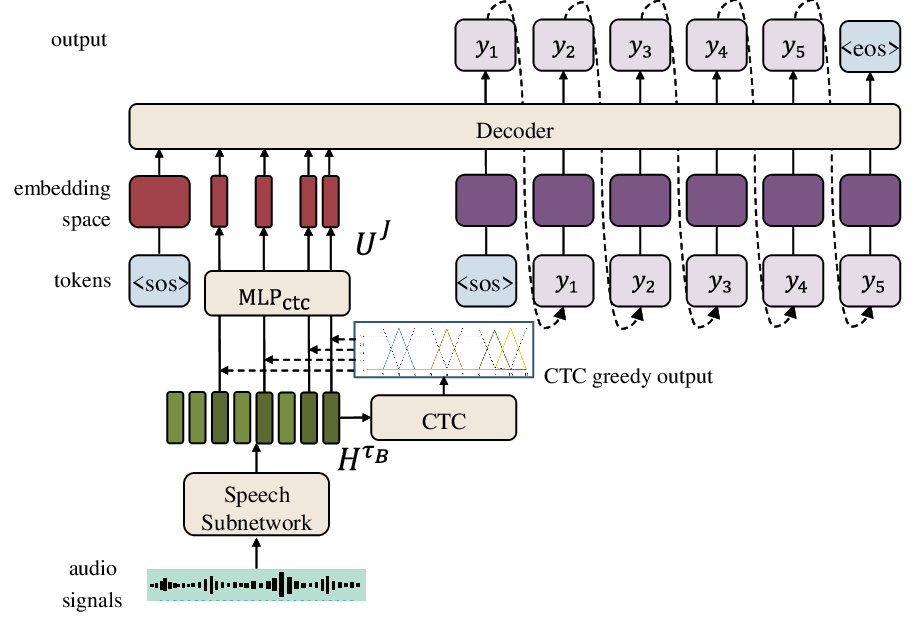}
  \vspace{-0.5cm}
  \caption{Decoder-only architecture for ASR. CTC greedy search outputs are used for filtering out speech frames to generate prompts for the decoder.}
  \label{fig:deconly}
\end{figure}

ASR is a task to predict the most probable $I$-length label sequence $\Y^I$ given a $T$-length input audio frame sequence $\X^{T}$.
In decoder-only architectures, instead of directly using the audio input $\X^{T}$ in the decoder, it is generally approximated by $J$-length compact audio representation $U^J$ as prompts ($J \ll T$) \cite{chang2023speechprompt,zhang2023speechgpt,rubenstein2023audiopalm,fathullah2023prompting,wu2023decoder,tsunoo2023decoder}. 
Prompts $U^J$ are generated using a speech subnetwork described in Sec.~\ref{ssec:blockproc} from the input $\X^{T}$.
The architecture is illustrated in the upper part of Fig.~\ref{fig:deconly}.
The compressed audio prompt $U^J$ is directly injected into the continuous embedding space of the decoder.
The decoder, then, autoregressively predicts the next linguistic token $y_i$ given the audio prompts $U^J$, and the previous outputs $\Y^{(i-1)}$ as follows:
\begin{align}
\vspace{-0.3cm}
    p_{\dec}(\Y^i) &= \prod_{k=1}^{i} p_{\dec}(y_k|U^J, \Y^{(k-1)}), \label{eq:dec}
\end{align}
where $y_0$ and $\U_0$ are a start-of-sequence token, $\langle$sos$\rangle$, and its embedding vector, respectively.

%\subsection{Block processing of prompt generation for streaming ASR with decoder-only architecture}
\subsection{Blockwise processing of prompt generation}
\label{ssec:blockproc}
In streaming ASR, blockwise processing is widely used \cite{dong19,tsunoo19, shi2021emformer}.
The speech subnetwork produces the prompts on-the-fly in each block.
%Let $B$ be the total block number, $T_b$ be the last frame of $b$-th block, and $J_b$ be the last prompt index corresponds to $X^{t \in b}=\{\x_t|T_{(b-1)} < t \leq T_b\}$.
%Chunks of prompts $U^{j\in b}=\{\U_j|J_{(b-1)} < j \leq J_b\}$ is generated by the speech subnetwork.
Let $B$ represent the total number of blocks, $T_b$ denote the last frame of the $b$-th block,% ($1\leq b \leq B$),
and $J_b$ signify the final index of the audio prompt for block $b$. 
Consequently, the chunk of audio prompt $U^{j\in b}=\{\U_j|J_{(b-1)} < j \leq J_b\}$ is generated by the speech subnetwork from speech input $X^{t \in b}=\{\x_t|T_{(b-1)} < t \leq T_b\}$ for the $b$-th block.
Thus, when the prompts are used for streaming ASR, the full audio prompt $U^J$ cannot be provided from the beginning.
Therefore, we propose a novel procedure that sequentially provides prompts in a blockwise manner.

\begin{figure}[t]
  \centering
  \includegraphics[width=0.8\linewidth]{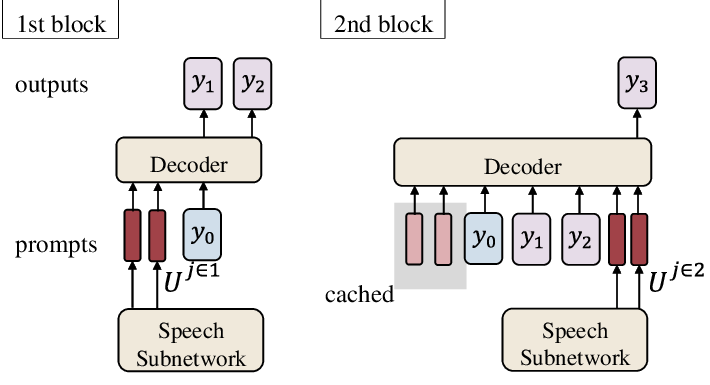}
  \vspace{-0.3cm}
  \caption{Blockwise processing of prompt generation for decoder-only architecture.}
  \label{fig:streaming}
\end{figure}

% At the $b$-th block, by applying Eq.~\eqref{eq:prompt}, the prompts are only created up to $\hat{\H}^{\tau_b}$ that corresponds to $\H^{T_b}$.
At the $b$-th block, the prompts are only created up to $U^{J_b}$.
The decoder estimates the output label sequence based on the partial prompts $U^{J_b}$ instead of $U^J$ in Eq.~\eqref{eq:dec}. 
\begin{align}
\vspace{-0.3cm}
    p_{\dec}(\Y^i) \approx \prod_{k=1}^{i} p_{\dec}(y_k|U^{J_b},\Y^{(k-1)}) \label{eq:streamingdec}
\vspace{-0.3cm}
\end{align}
Similarly in the $(b+1)$-th block, the additional prompts $U^{j\in (b+1)}$ is provided to the decoder, and the decoder continues estimation using $U^{J_{(b+1)}}$.
The series of computations is illustrated in Fig.~\ref{fig:streaming}.

The inference process is valid when the attention computations are properly masked out.
% When computing attentions for the added prompts, the mask is created so that they do not attend to the output tokens $Y^{(i-1)}$ and only attend to the past prompts $U^{J_{(b-1)}}$.
When computing attention for the added prompts $U^{j\in b}$, the mask is created such that they do not attend to the output tokens $Y^{(i-1)}$ and only attend to the past prompts $U^{J_{(b-1)}}$
% In addition, we use the absolute sinusoidal position encoding described in \cite{vaswani17} and individually apply it to prompts $\u_j$ and output tokens $y_i$.
% Therefore, both prompt $\hat{h}_i$ and token $y_i$ are applied with the same position encoding, regardless of whether they correspond each other or not.
% Therefore, $n^{th}$ token in audio prompt sequence ($\u_n$) and output token sequence ($y_n$) are applied with the same position encoding, regardless of whether they align with each other or not.
% Note that, as long as the corresponding position encoding is applied to the added prompts, the transformer decoder 
% can handle the prompts properly.
We also cache the prompts and computed intermediate values for efficiency, as highlighted on the right-hand side of Fig.~\ref{fig:streaming}.

\subsection{Prompts generation methods}
\label{ssec:prompts}
% The prompts are generated using the speech subnetwork and, since we compute them in a blockwise manner, we adopt contextual block processing \cite{tsunoo19} for the speech subnetwork.
As the prompts are computed in a blockwise manner using the speech subnetwork (see Sec.~\ref{ssec:blockproc}), we employ the contextual blockwise processing method \cite{tsunoo19} to construct our speech subnetwork.
%Our model is based on conformer \cite{gulati2020} and, even i
In blockwise processing, the model introduce an additional context embedding $\c_b$ to use not only local information within the $b$-th block but also global context information.
Each speech subnetwork layer inherits the context embedding $\c_b$ to the upper layer of the next block.
% Let the output of $n$-th layer of the speech subnetwork in the b-th block be $O_{n,b}$, then the speech subnetwork layer $S(\cdot)$ compute it based on the concatenated sequence of the previous layer output and the context embedding vector, as
Thus, the output at $n$-th layer of the speech subnetwork $S_n(\cdot)$ in the b-th block, $O_{n,b}$, is computed based on the concatenated sequence of the previous layer output $O_{(n-1),b}$ and the context embedding vector  $\c_{(n-1),(b-1)}$ as follows.
\begin{align}
\vspace{-0.3cm}
    O_{n,b} \oplus \c_{n,b} &= S_n(O_{(n-1),b} \oplus \c_{(n-1),(b-1)}), \label{eq:context} 
\vspace{-0.3cm}
\end{align}
where $\oplus$ is a concatenation operation.
% We first extract a subsampled acoustic representation sequence $\H^{T}=\{\h_t|1\leq t \leq T\}$ from input $\X^{T_0}$ using a speech subnetwork ($T_0 < T$).
% Finally, the output of the last layer is the subsampled acoustic representation sequence $\H^{t\in b}=\{\h_t|\tau_{b-1}< t \leq \tau_b\}$, which corresponds to the speech input $\X^{t\in b}$ ($\tau_B < T$).
Finally, the output of the last layer is considered as the subsampled acoustic representation sequence $\H^{t\in b}=\{\h_t|\tau_{b-1}< t \leq \tau_b\}$, which corresponds to the speech input $\X^{t\in b}$ ($\tau_b < T_b$), as 
\begin{align}
    H^{t\in b} &= O_{N,b}, \label{eq:representation}
\end{align}
where $N$ is the number of layers in the speech subnetwork.
%An acoustic representation sequence $\H^{t\in b}=\{\h_t|T_{b-1}< t \leq T_b\}$ is extracted as a by the speech subnetwork from a chunk of speech input $\X^{t\in b}$.
% the output of the last layer for $b$-th block $O_{N,b}$ is
By using this speech subnetwork, we propose two methods for generating prompts in the next subsections.

%Based on Eq.~\eqref{eq:prompt}, a chunk of prompts are provided from the $b$-th block.
%\begin{align}
%    \hat{\H}^{\tau \in b} = \{\MLP(\h_t)|t:\hat{z}_t\neq \phi,T_{b-1} < t \leq T_b\} \label{eq:partial}
% \end{align}
% \subsection{Decoder-only architecture for streaming ASR}
% \label{ssec:streaming}
% When the decoder-only architecture is combined with the blockwise processed described in Sec.~\ref{ssec:blockproc}, 
% Therefore, when the decoder-only architecture is combined with the blockwise processed speech subnetwork, 
% When the prompts are used for streaming ASR, the prompts $\hat{\H}^\tau$ cannot be provided in a full form from the beginning.
% Therefore, we propose a novel procedure that sequentially provides prompts in a blockwise manner.

\subsubsection{CTC prompts}
\label{sssec:ctcprompts}
% For the prompts, we propose two methods to compress 
Following \cite{wu2023decoder, tsunoo2023decoder,hono2023integration}, we use the greedy search output of CTC to efficiently generate the prompts.
% The speech subnetwork is similar to a normal speech encoder in \cite{karita19}, and w
We introduce CTC \cite{graves06, miao15, amodei16} with a blank token, $\phi$, to align $\H^{\tau_B}$ with the corresponding label sequence $\Y^{I}$.
CTC estimates $\phi$-augmented tokens $z_t \in \V \cup \{\phi\}$ at time frame $t$, where $\V$ is the token vocabulary.
% Based on $\h_{t}$, CTC calculates a posterior of $z_t$.
Based on $\h_{t}$ computed by the speech subnetwork (Eq.~\eqref{eq:representation}), CTC calculates a posterior of $z_t$.
\vspace{-0.1cm}
\begin{align}
    q(z_t|\h_{t}) = \CTC(\h_{t}) \label{eq:ctc}
\end{align}
The estimated sequence $\Z^{\tau_B}=\{z_t|1\leq t \leq \tau_B\}$ can be mapped to a label sequence using a mapping function $\F:\Z^{\tau_B}\rightarrow \Y^{I}$.

%To generate the prompts, CTC greedy search output $\hat{z}_t=\mathrm{arg}\max_{z_t} q(z_t|\h_t)$ filters out the frames that output blank token $\phi$, significantly reducing the sequence length.
% To generate the prompts, we 
We then perform CTC greedy search decoding to obtain $\hat{z}_t=\mathrm{arg}\max_{z_t} q(z_t|\h_t)$ and then selectively filter out the frames that output blank token $\phi$, thereby significantly reducing the sequence length.
% Thus, the audio information is provided as $\tau<T$-length prompts $\hat{\H}^{\tau} = \{\hat{\h}_{t}|0 \leq t < \tau\}$ directly in the embedding space of the decoder.
To obtain prompts $U^{j\in b}$ (See Sec.~\ref{ssec:blockproc}), the speech subnetwork outputs are mapped to the embedding space of the decoder with a linear layer ($\MLP_{\ctc}(\cdot)$) as:
\begin{align}
    U_{\ctc}^{j\in b} = \{\MLP_{\ctc}(\h_{t})|t:\tau_{(b-1)} < t \leq \tau_b, \hat{z}_t\neq \phi\}.\label{eq:ctcprompt}
\end{align}
The process is graphically shown in the lower part of Fig.~\ref{fig:deconly}.
%Therefore, in the case, prompt $\hat{h}_t$ corresponds to output $y_i$.
%In other words, to estimate $y_i$, the decoder may need $\hat{h}_t$ as a prompt for a acoustic clue.
% Therefore, $\U_n$ is thought to contain acoustic information corresponding to $y_n$.
%Note that, with the ideal CTC, the frames corresponding to CTC output $y_i$ remain for the prompts.
%Since the CTC output sometimes contains errors, the decoder not only performs a mere copy-and-paste of the prompts but also refines the CTC prediction with its language model capability.

\subsubsection{Block context prompts}
\label{sssec:context}
The context embedding vector in the last layer $\c_{N,b}$ in Eq.~\eqref{eq:context} considerably contains acoustic information of not only the $b$-th block but also of its context, i.e., previous blocks.
This contextual information can also be used as prompts to the decoder, in addition to the CTC prompts (Sec.~\ref{sssec:ctcprompts}).
We use another linear layer ($\MLP_{\ctx}(\cdot)$) to map the context embedding $\c_{N,b}$ to the embedding space of the decoder.%, as shown in Fig.~\ref{fig:context}.
\vspace{-0.1cm}
\begin{align}
    U_{\ctx}^{j\in b} = \MLP_{\ctx}(\c_{N,b}) \label{eq:contextprompt}
\end{align}
Thus, we concatenate $U_{\ctc}^{j \in b}$ in Eq.~\eqref{eq:ctcprompt} and $U_{\ctx}^{j\in b}$ in Eq.~\eqref{eq:contextprompt} to obtain prompts for the decoder $U^{j \in b}$ (See Eq.~\eqref{eq:streamingdec}). %prompts for the decoder, as
\vspace{-0.1cm}
\begin{align}
    U^{j \in b} = U_{\ctc}^{j \in b} \oplus U_{\ctx}^{j\in b} \label{eq:prompt}
\end{align}

\begin{comment}
\begin{figure}[t]
  \centering
  \includegraphics[width=0.9\linewidth]{context2.eps}
  \caption{Context embedding of each block from the last layer of the speech subnetwork is additionally incorporated as prompts to the decoder.}
  \label{fig:context}
\end{figure}
\end{comment}

\subsection{Score fusion inference}
\label{ssec:inference}
The probabilistic scores of CTC and the decoder are combined either in a frame-synchronous manner \cite{graves06} or in a label-synchronous manner \cite{watanabe17}.
In particular, for streaming ASR, the integrated variation of frame/label-synchronous beam search improves performance \cite{tsunoo23_interspeech}.
We propose using this integrated beam search for the streaming decoder-only architecture.

In the beam search both hypotheses generated by CTC and the decoder are kept for score fusion.
The score is computed based on the log-likelihood of both CTC probability $p_{\ctc}(\Y_{\ctc}^l,t)$ and decoder probability $p_{\dec}(\Y_{\dec}^i)$ (Eq.~\eqref{eq:streamingdec}), where $\Y_{\ctc}^l$ and $\Y_{\dec}^i$ are $l$-length hypothesis based on the frame-synchronous search of CTC and $i$-length hypothesis based on the label-synchronous search of the decoder, respectively.
Note that, according to \cite{tsunoo23_interspeech}, $i$ is upper bounded by $l$ as the decoder hypotheses $\Y_{\dec}^i$ is always a prefix of the CTC hypotheses $\Y_{\ctc}^l$.
$p_{\ctc}(\Y_{\ctc}^l,t)$ is efficiently calculated by recursively accumulating the probability $\gamma(\cdot)$, based on the frame-wise posteriors $q(z_t|\h_t)$ (Eq.~\eqref{eq:ctc}), of the sequence ending with blank $\Y^l_\b$ and non-blank $\Y^l_\n$ as 
\vspace{-0.1cm}
\begin{align}
    p_{\ctc}(\Y_{\ctc}^{l},t) &= \gamma(\Y_\b^{l},t) + \gamma(\Y_\n^{l},t) \label{eq:fsprob} \\
    \gamma(\Y_\b^{l},t) &= \sum_{\F_{\ctc}(\Z^{t})=\Y^{l}:z_t=\phi}p_{\ctc}(\Y_{\ctc}^{l}|\H^{t-1})q(z_t|\h_t) \nonumber \\
    \gamma(\Y_\n^{l},t) &= \sum_{\F_{\ctc}(\Z^{t})=\Y^{l}:z_t=y_l}p_{\ctc}(\Y_{\ctc}^{l-1}|\H^{t-1})q(z_t|\h_t). \nonumber 
\end{align}
The score fusion is done in a frame/label two-dimension grid, $(t,i)$ based on log-likelihood of $p_{\dec} (\Y_{\dec}^i)$ in Eq.~\eqref{eq:streamingdec} and $p_{\ctc}(Y_{\ctc}^l,t)$ in Eq.~\eqref{eq:fsprob}, as
\vspace{-0.1cm}
\begin{align}
    s(\Y^l, t, i) = \lambda_{\ctc} \log(p_{\ctc}(Y_{\ctc}^l,t)) + \lambda_{\dec} \log( p_{\dec} (\Y_{\dec}^i)), \label{eq:score}
\end{align}
where $\lambda_{*}$ are fusion weights.
LMs and length penalty can also be combined while scoring the hypothesis as in \cite{tsunoo23_interspeech}.

\section{Training procedure}
\label{sec:training}
\subsection{Fine-tuning models}
\label{ssec:finetuning}
Training all the parameters of the decoder-only architecture from scratch is not effective on a large scale according to our previous work \cite{tsunoo2023decoder}.
Therefore, we follow the fine-tuning strategies described in \cite{fathullah2023prompting,wu2023decoder} for the proposed streaming ASR model.
We first train the blockwise speech subnetwork with CTC based on \cite{tsunoo19} and a decoder-only transformer LM individually.
Subsequently, we combine each module to perform fine-tuning.
In each training step, the CTC greedy output $\hat{z}_t$ (Sec.~\ref{sssec:ctcprompts}) is computed and the prompts $U^{j\in b}$ in Eq.~\eqref{eq:prompt} are generated on-the-fly for each block $b$.
Using the prompts, the decoder is trained to minimize a cross entropy loss, and the parameters of the speech subnetwork are also updated.
%We conduct an experiment to compare training from scratch and the fine-tuning approach in Sec.~\ref{ssec:scale}.

\subsection{Prompt masking in training}
\label{ssec:promptmasking}
% As discussed in Sec.~\ref{ssec:deconly}, ideally prompt $\hat{h}_i$ corresponds to output $y_i$.
\subsubsection{Full prompts}
\label{sssec:fullprompts}

For streaming inference, to estimate $y_i$ at $b$-th block, only a partial of prompts $U^{J_b}$ are provided.
However, it is not trivial to prepare partial transcription corresponding to the prompts without accurate forced alignment.
The naive training using the entire prompts $U^J$ for the transcription $Y^I$ may cause training--inference mismatch, i.e., Eq.~\eqref{eq:dec} is calculated in training while Eq.~\eqref{eq:streamingdec} is calculated in the inference step.
%the prompts are sometimes limited and a mismatch may occur when the decoder is always trained using entire prompts $\hat{\H}^{\tau_B}$ based on Eq.~\eqref{eq:dec}.
%However, during training, it is not trivial to align $\hat{\H}^{\tau_B}$ and $\Y^I$.
%In this work, we propose three methods to train the model to mitigate the mismatch between training and streaming inference.
Therefore, we propose two methods to mask out the prompts while training to mitigate the mismatch.

% The first method is simply training using entire prompts $\hat{\H}^{\tau_B}$ following Eq.~\eqref{eq:dec}.
% This causes the mismatch and we see how much this mismatch affects to the performance in Sec.~\ref{ssec:promptexp}.

\subsubsection{Forced alignment masking}
\label{sssec:forcedalign}
To strictly align $U^J$ and $\Y^I$, we propose using CTC forced alignment \cite{kurzinger2020ctc} to mask out future information.
The forced alignment $a$ aligns output index $i(a)$ with frame $\tau(a)$.
% When computing the attention weights for $y_{i(a)}$, we compute the corresponding prompts as follows.
When the decoder estimates $y_{i(a)}$, we compute the corresponding prompts as follows.
\begin{align}
    U_{\ctc}^{y_{i(a)}} &= \{\MLP_{\ctc}(\h_{t})|t:0 < t \leq \tau(a), \hat{z}_t\neq \phi\} \\
    U_{\ctx}^{y_{i(a)}} &= \{\MLP_{\ctx}(\c_{N,b})|b:\tau_{b-1} \leq \tau(a)\} \label{eq:forcedalign}
\end{align}
This calculation is efficiently done by manipulating the mask for self-attention computation.
The mask is created on-the-fly using CTC of the latest parameters.

\begin{table*}[t]
\caption{Performance comparison for streaming ASR trained with LibriSpeech.}
\vspace{-0.3cm}
\label{tab:lib960}
    \centering
    % \hspace{-0.3cm}
    \begin{tabular}{l|c|c|cc|cc|c|c}
     \hline
    Models & \# of  &Training &  \multicolumn{2}{c|}{Batched WER} & \multicolumn{2}{c|}{Streaming WER} & RTF & Latency \\
    & pram.&method&test-clean & test-other &test-clean & test-other && EP50 (sec) \\
    \hline\hline
         % Baseline CTC \cite{graves06} &&  4.9&12.4& 4.9 & 12.4 & 0.07 & 0.09\\ w/o LM
         Baseline CTC \cite{graves06} &41.5M&&  3.9&\hphantom{0}9.7& 3.9 & \hphantom{0}9.7 & 0.07 & 0.09\\
         Baseline RNNT \cite{graves13rnnt}&43.5M& &3.5 & \hphantom{0}9.5 & 3.5 & \hphantom{0}9.5 & 0.08 & 0.15 \\ % w/o LM
         % Baseline EncDec \cite{tsunoo23_interspeech}& &3.8 & \hphantom{0}9.4 & 3.7 & \hphantom{0}9.8 & 0.37 & 0.92 \\ w/o LM
         Baseline EncDec \cite{tsunoo23_interspeech}&53.0M& &3.5 & \hphantom{0}8.4 & {\bf 3.2} & \hphantom{0}8.6 & 0.37 & 0.92 \\
         \hline
         Decoder-only (proposed) &&&&&&&&\\  
         % CTC prompts (Eq.~\eqref{eq:ctcprompt}) & full prompt & 3.6 & 9.3 & (3.7) & (9.6) & & \\ %w/o LM
         CTC prompts (Eq.~\eqref{eq:ctcprompt}) &51.5M& full prompt & 3.3 & \hphantom{0}8.7 & 3.5 & \hphantom{0}9.0 &0.17 & 0.37\\ %w/o LM
         &&forced align& 7.8 & 15.1 & 8.3 & 14.7 &0.12 &0.36 \\ % w/o LM
         % &prefix&(5.7) & (10.0) & 3.6 & \hphantom{0}8.8 & 0.17 & 0.44\\ w/o LM
         &&prefix&3.5 & \hphantom{0}8.2 & 3.6 & \hphantom{0}8.0 & 0.17 & 0.44\\
         % Context prompts (Eq.~\eqref{eq:contextprompt}) &prefix& 3.7 & 9.1 &  &  & 0.12 & 0.37\\ w/o LM
         Context prompts (Eq.~\eqref{eq:contextprompt}) &51.5M&prefix& 4.2 &\hphantom{0}9.0 & 3.8 & \hphantom{0}9.0 & 0.12 & 0.37\\
         % CTC+Context prompts (Eq.~\eqref{eq:prompt}) &prefix &4.0&9.3& & & 0.16 & 0.47\\ w/o LM
         CTC+Context prompts (Eq.~\eqref{eq:prompt}) &51.5M&prefix &3.4&\hphantom{0}8.0& {\bf 3.2}& {\bf \hphantom{0}7.9} & 0.16 & 0.47\\          
         \hline
    \end{tabular}
      \vspace{-0.5cm}
\end{table*}

\subsubsection{Prefix prompts}
\label{sssec:prefix}
Instead of strictly aligning prompts and the output labels, we also propose randomly selecting a prefix of the prompts.
The random block size $1\leq \beta \leq B$ is selected at each training step, and the prefix prompts $U^{J_\beta}$ is used to estimate the entire sentence $Y^I$ using Eq.~\eqref{eq:streamingdec}.
% Therefore, in the ideal case of CTC working perfectly, only $y_{\leq r}$ has sufficient acoustic clue, and the following $y_{> r}$ cannot depend on the acoustic information.
Because it does not rely on the CTC module, we expect this method to be more robust to the CTC error.
%We regard this training method as a mixture of ASR/LM training, because $y_i$ corresponding to $X^{T_\beta}$ uses sufficient acoustic clue of $U^{J_\beta}$, while the rest estimation is regarded as pure LM training as it practically only depends on the previous outputs, i.e., $p(y_i|\Y^{(i-1)})$.
We consider this training method similar to a mixture of ASR/LM training. 
This is because the tokens $y_i$ are aligned with $\x_{t}$ such that $t<T_\beta$ benefit from sufficient acoustic clues from $U^{J_\beta}$, resembling ASR training. 
Conversely, the estimation for the remaining tokens $y_i$ can be regarded as pure LM training as it primarily relies on previous outputs, i.e., $p(y_i|\Y^{(i-1)})$, rather than Eq.~\eqref{eq:streamingdec}.

\section{Experiments}
\label{sec:experiments}
%\subsection{Training decoder-only architecture at scale}
%\label{ssec:scale}
\subsection{Experimental setup}
\label{ssec:setup}
%First, we confirmed whether the fine-tuning approach was more effective than training from scratch in a large corpus of LibriSpeech \cite{panayotov15}, as discussed in Sec.~\ref{ssec:finetuning}.
%For this experiment, we carried out batch-processing for inference to see a pure behavior of a decoder-only model.
To evaluate the proposed decoder-only architecture for streaming ASR, we used LibriSpeech \cite{panayotov15} and Switchboard dataset.
We first trained the speech subnetwork and the decoder, which was a 12-block conformer and 6-block transformer, respectively, both with four-head 256-unit attention layers and 2048-unit feed-forward layers, and the other detail including block size were as in \cite{tsunoo23_interspeech}.
The speech subnetwork was trained with the CTC loss. %using a weight of $\lambda=0.3$.
% For the decoder-only architecture, the transformer with the same configuration was pretrained using the external text-only corpus.
For the decoder pre-training, we used the external text-only corpus for LibriSpeech, and Fisher corpus for Switchboard.
We used the Adam optimizer with a learning rate of 0.0025 decayed by Noam learning rate decay.
Then the entire model consisting of both speech subnetwork and decoder-only architecture was fine-tuned on paired speech--transcription data using a lower learning rate of 0.0005 without warm-up.
%As a comparison, we also trained the model form scratch following \cite{tsunoo2023decoder}.
The external LMs, used for scoring hypotheses during inference, were also trained using text-only data. %, including an additional text corpus from LibriSpeech. 
The LMs were two-layer LSTMs with 512 hidden units.
We applied byte-pair encoding subword tokenization with 5,000 token classes for LibriSpeech and 2,000 token classes for Switchboard, respectively.
% $\lambda_{\ctc}=0.4$ performed best in the development set, and the LM was also fused with a weight of 0.4.

For comparison, we also trained the baseline conformer-based encoder--decoder streaming model \cite{tsunoo23_interspeech} and used it as Baseline EncDec or Baseline CTC by omitting the decoder.
We also added RNN transducer layers \cite{graves13rnnt} to the same encoder architecture as another baseline (Baseline RNNT).
The baselines methods were decoded using the integrated frame/label-synchronous search \cite{tsunoo23_interspeech} for EncDec and frame-synchronous search \cite{graves06} for the rest.
The fusing weights in Eq.~\eqref{eq:score} were determined using the development set, and were $\lambda_{\ctc}=0.4$ and $\lambda_{\dec}=0.6$.
The LM was also fused with a weight of 0.4.

We measured not only word error rates (WERs), but also real-time factors (RTFs) and latency using randomly selected 100 utterances from the test set of each corpus with an 8 core 3.60 GHz Intel i9-9900K CPU.
We adopted EP latency following \cite{yu2021fastemit}, which was the duration between the last audio sample inputted and the last transcription token emitted.
The median of both RTFs and latency are reported.

\begin{comment}
\begin{table}[t]
\caption{Non-streaming ASR results with LibriSpeech 960h. The external LM was fused to all the models.}

\label{tab:16block}
    \centering
    \begin{tabular}{l|cc}
     \hline
    Models & \multicolumn{2}{c}{WER}\\
    & test-clean & test-other  \\
    \hline\hline
         Baseline EncDec \cite{guo2020recent}&  2.8 & 6.4 \\
         \hline
         Decoder-only && \\
         \ \ \ {\it from scratch} \cite{tsunoo2023decoder} & 2.8 & 6.6 \\
         \ \ \ {\it fine-tuning} (Sec.~\ref{ssec:finetuning}) & {\bf 2.4} & {\bf 5.7} \\
         \hline
    \end{tabular}
      \vspace{-0.4cm}
\end{table}

The results are shown in Table.~\ref{tab:16block}.
For the large scale as Librispeech corpus, training from scratch was not effective as \cite{tsunoo2023decoder} reported.
However, the fine-tuning method exploit thoroughly studied LM training and efficiently adapted to the ASR task.
In the following experiments, we adopted the fine-tuning approach for our proposed streaming decoder-only model as in Sec.~\ref{ssec:finetuning}.
\end{comment}

% \subsection{Comparison of prompt masking methods}
% \label{ssec:promptexp}
\subsection{Librispeech results}
\label{ssec:librispeech}
We investigate the efficacy of decoder-only architecture for streaming ASR using LibriSpeech.  
All the methods were inferred not only in a streaming manner (Sec.~\ref{ssec:blockproc}), but we also performed an ablation study of batch processing inference.
In the case of batch processing, the full prompts were given to the decoder, which was a matched condition with full prompt training using Eq.~\eqref{eq:dec}, described in Sec.~\ref{sssec:fullprompts}.
% We trained the models using LibriSpeech \cite{panayotov15} as in Sec.~\ref{ssec:scale}.
% The model training configuration followed Sec.~\ref{ssec:scale} except that the number of block was 6 for the decoder.
% We omitted LM fusion for computational efficiency and a pure comparison of the models themselves.

First, we conducted experiments with the three training methods discussed in Sec.~\ref{ssec:promptmasking}, using only CTC prompts, i.e., $U^J=U^J_{\ctc}$ (Eq.~\eqref{eq:ctcprompt}).
The results are shown in Table~\ref{tab:lib960}.
For the naive full prompt model (Sec.~\ref{sssec:fullprompts}), the inference using full prompts is a matched condition. %, and it performed the best among all the other decoder-only methods.
However, when it was inferred using the proposed streaming processing, the train--inference mismatch (see Sec.~\ref{sssec:fullprompts}) degraded the accuracy, from 3.3\% to 3.5\% and 8.7\% to 9.0\% for each test set.
For the forced alignment model (Sec.~\ref{sssec:forcedalign}), the model did not perform well and the training was unstable.
We hypothesize that this was because the training used different alignments given by new CTC parameters in every iteration and the CTC forced alignment might contain errors.
On the other hand, training using the random-length prefix prompts (Sec.~\ref{sssec:prefix}) performed robustly in both the mismatched batched scenario and the target streaming scenario.
In particular, the prefix prompt training reduced test-other WERs significantly, i.e., by 0.5\% and 1.0\% in both the batched and streaming inference, compared to full prompt training.
The result indicates the efficacy of prefix prompt training in covering various situations of prompting that the decoder faces in inference.
Among the three training methods, prefix prompt training scored lowest WER in the streaming scenario, and outperformed Baseline CTC/RNNT models.% , which was comparable to the Baseline EncDec.
%Comparing to the naive training using full prompts, the prefix prompt training performed better in 1.4\% and 2.5\% absolute in test-clean and test-other sets with the streaming inference.
%In the following experiments, we trained our proposed decoder-only streaming ASR models with prefix prompt method.

\begin{comment}
\begin{table*}[t]
\caption{Training method comparison for streaming ASR trained with LibriSpeech 960h.}
\label{tab:training}
    \centering
    \begin{tabular}{l|cc|cc}
     \hline
    Models--{\it training} & \multicolumn{2}{c|}{batch WER} & \multicolumn{2}{c}{streaming WER} \\
    &test-clean & test-other & test-clean & test-other  \\
    \hline\hline
          Full prompts (Sec.\ref{sssec:fullprompts}) & 3.6 &  \hphantom{0}9.3 & 5.8 & 12.1 \\
         Forced align (Sec.\ref{sssec:forcedalign}) & 7.1 & 13.5 & 8.3 & 13.8 \\
         Prefix (Sec.\ref{sssec:prefix}) & 5.7 & 10.0 &4.4 & \hphantom{0}9.6 \\
         \hline
    \end{tabular}
      \vspace{-0.4cm}
\end{table*}

\subsection{Performance comparison}
\end{comment}

%For this experiments, we also added the block context prompts $U^J_{\ctx}$ in Eq.~\ref{eq:contextprompt}.
% For the proposed decoder-only architecture, 
Then, we evaluated the additional block context prompts $U^J_{\ctx}$ in Eq.~\eqref{eq:contextprompt}.
As shown in Table~\ref{tab:lib960}, using only the block context prompts performed 3.8\% and 9.0\% for test-clean/other sets, respectively.
When we combined two proposed prompts as $U^J$ (Eq.~\eqref{eq:prompt}), we observed improvement over CTC prompt model by 0.4\% and 0.1\% absolute, while maintaining nearly the same computational efficiency, with latency increasing by only 0.03 seconds.
Compared to Baseline EncDec, the CTC+Context prompt model achieved similar performance in test-clean and 8\% relative WER reduction in test-other with almost 50\% reduction in computational requirements.

\begin{table}[t]
\caption{Streaming ASR evaluation using Switchboard.}
\vspace{-0.3cm}
\label{tab:swbd}
%    \centering
\hspace{-0.2cm}
    \begin{tabular}{l|cc|c}
     \hline
    Models & \multicolumn{2}{c|}{Streaming WER} & RTF\\
    & Switchboard & CallHome  &\\
    \hline\hline
         Baseline CTC \cite{graves06} & 10.2& 17.8 &0.06\\
         Baseline RNNT \cite{graves13rnnt}&10.3 & 17.7 &0.09\\
         Baseline EncDec \cite{tsunoo23_interspeech}& \hphantom{0}8.8 & 16.0 &0.46\\
         \hline
         Decoder-only (Eq.~\eqref{eq:prompt}) &{\bf \hphantom{0}8.7} & {\bf 15.7} &0.39\\
         \hline
    \end{tabular}
      \vspace{-0.5cm}
\end{table}

\begin{comment}

\begin{table}[t]
\caption{Streaming ASR evaluation using Switchboard.}
\label{tab:swbd}
    \centering
    \begin{tabular}{l|cc}
     \hline
    Models & \multicolumn{2}{c}{WER} \\
    & Switchboard & CallHome  \\
    \hline\hline
         Baseline CTC \cite{graves06} & 10.9& 18.7 \\
         Baseline RNNT \cite{graves13rnnt}&10.3 & 17.7 \\
         Baseline EncDec \cite{tsunoo23_interspeech}& 9.2 & 16.3 \\
         \hline
         Decoder-only (Eq.~\eqref{eq:prompt}) &{\bf 9.0} & {\bf 16.2} \\
         \hline
    \end{tabular}
      \vspace{-0.4cm}
\end{table}
\end{comment}

\subsection{Switchboard and Fisher results}
\label{ssec:swbd}
Lastly, we evaluated the proposed method trained with the Switchboard dataset to see if our method is consistently effective in a different dataset.
% Table~\ref{tab:swbd} shows the results.
The results in Table.~\ref{tab:swbd} follows similar tendency as in Librispeech corpus, and we confirmed that our proposed methods outperformed all baselines by achieving WERs of 9.0\% and 16.2\% for Switchboard/CallHome test set.

\section{Conclusion}
\label{sec:conclusion}
We investigated using powerful yet efficient decoder-only architecture for blockwise streaming ASR.
We propose a novel architecture where the speech utterances are processed in a blockwise speech subnetwork, and prompts are produced block-by-block based on CTC greedy output and the context embedding. %, which represent acoustic information.
The decoder-only architecture estimates transcriptions in a blockwise manner after each chunk of prompts are given.
For training, we also propose novel \emph{prefix prompt training} that mitigate the mismatch between batched training and streaming inference.
This method selects the prefix of the prompts randomly to cover various prompt lengths that may occur in inference.
Experimentally we confirmed that our proposed usage of decoder-only architecture in streaming ASR performs better accuracy with smaller RTF and latency. %, e.g., 0.7\% WER reduction on Librispeech test-other set with almost 50\% less computational cost.
\bibliographystyle{IEEEtran}
\bibliography{mybib}

\end{document}